\documentclass[conference]{IEEEtran}
\IEEEoverridecommandlockouts
\usepackage{amsmath,amssymb,amsfonts}
\usepackage{algorithmic}
\usepackage{cite}
\usepackage{textcomp}
\usepackage{listings}
\usepackage{xcolor}
\usepackage{graphicx}
\usepackage{graphics}
\usepackage[caption=false,font=footnotesize]{subfig}
\usepackage{balance}
\usepackage{comment}
\usepackage[ruled,linesnumbered]{algorithm2e}
\usepackage{multirow}
\usepackage{makecell}
\usepackage{float}
\usepackage{hyperref}

\def\BibTeX{{\rm B\kern-.05em{\sc i\kern-.025em b}\kern-.08em
    T\kern-.1667em\lower.7ex\hbox{E}\kern-.125emX}}
    
\definecolor{mygreen}{rgb}{0,0.6,0}
\definecolor{mygray}{rgb}{0.5,0.5,0.5}
\definecolor{mymauve}{rgb}{0.58,0,0.82}

\begin{document}

\title{Accelerating Post-Quantum Cryptography via LLM-Driven Hardware-Software Co-Design}

\author{\IEEEauthorblockN{Yuchao Liao}
\IEEEauthorblockA{\textit{Electrical \& Computer Engineering} \\
\textit{University of Arizona}\\
Tucson, AZ, USA \\
yuchaoliao@arizona.edu}
\and
\IEEEauthorblockN{Tosiron Adegbija}
\IEEEauthorblockA{\textit{Electrical \& Computer Engineering} \\
\textit{University of Arizona}\\
Tucson, AZ, USA \\
tosiron@arizona.edu}
\and
\IEEEauthorblockN{Roman Lysecky}
\IEEEauthorblockA{\textit{Electrical \& Computer Engineering} \\
\textit{University of Arizona}\\
Tucson, AZ, USA \\
rlysecky@arizona.edu}
}

\maketitle

\begin{abstract}
Post-quantum cryptography (PQC) is crucial for securing data against emerging quantum threats. However, its algorithms are computationally complex and difficult to implement efficiently on hardware. In this paper, we explore the potential of Large Language Models (LLMs) to accelerate the hardware-software co-design process for PQC, with a focus on the FALCON digital signature scheme. We present a novel framework that leverages LLMs to analyze PQC algorithms, identify performance-critical components, and generate candidate hardware descriptions for FPGA implementation. We present the first quantitative comparison between LLM-driven synthesis and conventional HLS-based approaches for low-level compute-intensive kernels in FALCON, showing that human-in-the-loop LLM-generated accelerators can achieve up to $2.6\times$ speedup in kernel execution time with shorter critical paths, while highlighting trade-offs in resource utilization and power consumption. Our results suggest that LLMs can minimize design effort and development time by automating FPGA accelerator design iterations for PQC algorithms, offering a promising new direction for rapid and adaptive PQC accelerator design on FPGAs.
\end{abstract}


\date{August 2025}
\maketitle

\section{Introduction}\label{sec:intro}
The advent of quantum computing presents both unprecedented opportunities and existential challenges to the field of cryptography. Quantum algorithms such as Shor's \cite{monz2016realization} and Grover's \cite{grassl2016applying} threaten to break widely deployed public-key cryptosystems, potentially compromising the confidentiality and integrity of sensitive data, digital signatures, and secure communications. To address this looming threat, the cryptography community has been actively developing post-quantum cryptography (PQC), a family of cryptographic algorithms designed to withstand attacks from both classical and quantum computers. The urgency of this transition is underscored by ongoing standardization efforts led by the National Institute of Standards and Technology (NIST), which has already selected candidate algorithms for both encryption and digital signatures.

Building on this urgency, NIST has spearheaded global efforts to standardize PQC algorithms. Among the leading candidates, the FALCON---Fast-Fourier Lattice-based Compact Signatures over NTRU (N-th degree Truncated Polynomial Ring) algorithm stands out for its ability to deliver compact signatures and efficient verification, making it particularly attractive for applications with strict bandwidth and storage constraints \cite{nistPQC2024}. Despite these advantages, however, like many PQC algorithms, FALCON's implementation remains challenging due to its reliance on complex mathematical operations and substantial resource requirements. These challenges motivate the exploration of hardware acceleration, particularly on FPGAs, which offer a reconfigurable platform capable of balancing high performance, energy efficiency, and adaptability to evolving cryptographic standards. In practice, different FALCON parameter sets (for example FALCON-512 and FALCON-1024) and evolving PQC standard revisions require hardware redesign or non-trivial adaptation, which makes rapid and repeatable design cycles particularly valuable for PQC accelerators.

The hardware-software co-design process for PQC algorithms such as FALCON has traditionally relied on a time-consuming, iterative workflow \cite{lee2024efficient,nguyen2020high}. Achieving efficient implementations demands deep expertise in both cryptography and hardware architecture, as designers must navigate an enormous design space while balancing performance, resource utilization, and security constraints. Prior efforts have primarily focused on using Register Transfer Level (RTL) design methodologies, which, while capable of delivering high performance, often require extensive manual optimization and lengthy development cycles \cite{oder2019_keyExch_FPGA,kuo2017high}. Within this context, hardware implementations of the Number Theoretic Transform (NTT)---a core component of many lattice-based PQC schemes---initially targeted Ring-LWE systems \cite{renteria2017high}, before later being tailored to specific PQC candidates like NewHope \cite{oder2019_keyExch_FPGA,nguyen2020high}. Although these approaches have proven effective, they highlight a persistent challenge: the bottleneck of PQC acceleration is dominated by the prohibitive cost in time, expertise, and manual effort required to produce high-quality FPGA designs.

In recent years, Large Language Models (LLMs) have demonstrated transformative potential in various domains that demand reasoning, code generation, and design space exploration \cite{liao2024llmsgoodhighlevelsynthesis}. LLMs are trained on vast amounts of data and can synthesize context-aware outputs that capture both high-level abstractions and low-level implementation details. The potential of LLMs to assist in technical tasks has been explored in fields like software engineering, system-level design, and even high-level synthesis (HLS), where LLMs have been shown to accelerate code development and automate optimization strategies \cite{collini2024c2hlscllmsbridgesoftwaretohardware,thakur2023verigenlargelanguagemodel}. Yet, despite the growing adoption of PQC and the urgent demand for efficient hardware support, the application of LLMs to cryptographic hardware-software co-design remains largely unexplored.

In this work, we explore a novel framework that leverages LLMs to accelerate the development and optimization of PQC implementations on FPGAs, with a specific focus on the FALCON algorithm. Our approach integrates LLM-driven analysis to automatically identify computational bottlenecks within the FALCON algorithm, determine components most suitable for hardware acceleration, and generate candidate hardware description language (HDL) implementations. By bridging algorithmic profiling with LLM-driven synthesis, this framework has the potential to drastically reduce design time, lower the barrier to entry for cryptographic hardware development, and open new directions in rapid, adaptive PQC accelerator generation.

\noindent \textbf{Summary of Contributions.} To the best of our knowledge, this is the first work to explore the use of Large Language Models (LLMs) for post-quantum cryptographic hardware acceleration on FPGAs. The key contributions of this paper are as follows:  
\begin{enumerate}
    \item \textbf{LLM-driven co-design framework:} We introduce a framework that leverages LLMs to guide hardware-software co-design of post-quantum cryptographic algorithms, with a focus on FPGA acceleration.
    \item \textbf{Application to FALCON:} We demonstrate the framework's applicability using the FALCON digital signature scheme by analyzing its algorithmic structure and identifying bottlenecks amenable to FPGA acceleration.  
    \item \textbf{Evaluation of automated HDL generation and optimization approaches:} We compare LLM-driven synthesis with conventional HLS-based methods for generating and refining hardware modules, highlighting their effectiveness in reducing reliance on manual RTL development and hardware-specific expertise.    
    \item \textbf{Evaluation and comparative insights:} Through place-and-route analysis of low-level compute-intensive functions, we show that LLM-generated accelerators achieve up to $2.6\times$ reduction in execution time and shorter critical paths compared to HLS-generated designs. These improvements come with trade-offs in resource utilization (LUTs/FFs) and power, highlighting the complementary strengths of LLM- and HLS-based approaches for PQC acceleration.    
  
\end{enumerate}

\begin{figure*}[th!]
\centering
\includegraphics[width=0.85\textwidth,keepaspectratio]{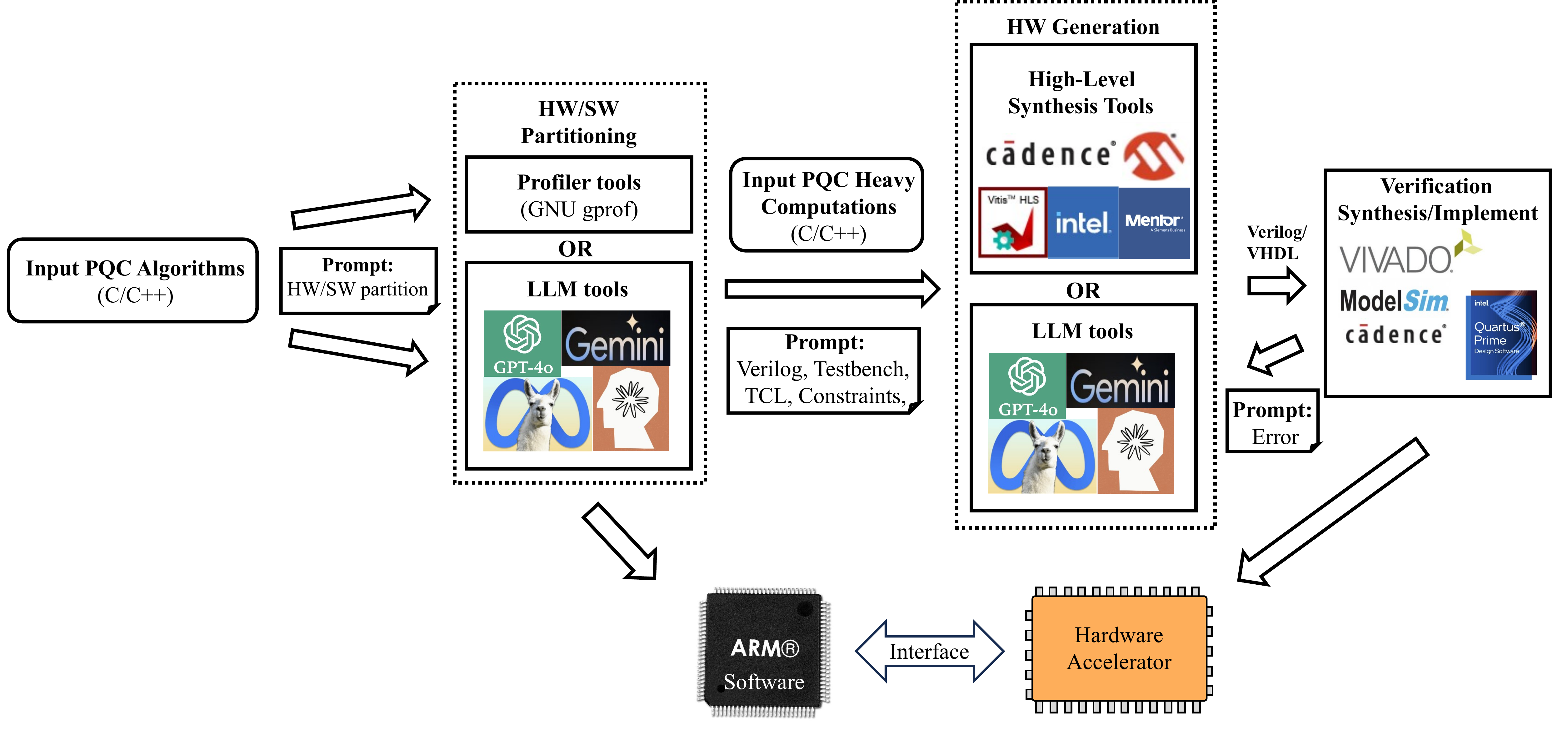}
\caption{Overview of the proposed LLM-driven PQC acceleration flow, integrating profiling, partitioning, and Verilog generation within FPGA design tools for rapid HW–SW co-design.}
\label{fig:methodology}
\vspace{-10pt}
\end{figure*}

\section{Background}\label{sec:relatedwork}
The emergence of practical quantum computers poses a critical threat to widely deployed cryptographic systems, particularly those relying on integer factorization (RSA) and discrete logarithmic problems (ECC, Diffie--Hellman) \cite{gitonga2025impact}. Post-quantum cryptography (PQC) aims to develop cryptographic algorithms that remain secure against both classical and quantum attacks. To prepare for the transition, the National Institute of Standards and Technology (NIST) initiated a multi-round standardization process, which has culminated in the selection of several lattice-based schemes as future standards. Among the standardization candidates, FALCON (Fast Fourier Lattice-based Compact Signatures Over NTRU) has been selected as a standard for digital signatures \cite{nistPQC2024}. FALCON belongs to the family of lattice-based cryptography schemes and is built on the NTRU (N-th degree Truncated Polynomial Ring) lattice structure, offering compact signatures and fast verification.

\subsection{FALCON and hardware acceleration}
FALCON's computational core can involve several mathematically intensive operations, including Number Theoretic Transforms (NTT), Fast Fourier Transforms (FFT) over the complex field, discrete Gaussian sampling, and NTRU polynomial arithmetic. While its compact signature size and verification efficiency make it appealing for bandwidth- and storage-constrained systems, the cost of signature and key generation remains high, dominated by large polynomial multiplications and transforms \cite{kim2022_FALCON_ARM}. These operations require careful modular arithmetic, floating-point precision handling, and secure randomness, all of which introduce implementation challenges.

Hardware acceleration \cite{karl2025high, kostalabros2021hls, kuo2017high, oder2019_keyExch_FPGA} thus becomes an attractive strategy to reduce latency and improve throughput. On FPGAs, custom datapaths can exploit fine-grained parallelism in polynomial arithmetic and FFT/NTT computations, while memory hierarchies can be tuned to handle large polynomial data efficiently. Moreover, FPGA reconfigurability allows support for multiple security levels and parameter sets, making them a compelling platform for PQC deployment. However, manually crafting RTL for such complex, multi-stage algorithms demands expertise and extended design cycles.  

\subsection{High-Level synthesis and LLM-assisted hardware design}
Conventional approaches to PQC hardware acceleration typically fall into two categories: (i) manual hand-optimized RTL design, or (ii) high-level synthesis (HLS) from C/C++ implementations \cite{kostalabros2021hls}. While RTL designs can achieve near-optimal performance and area utilization, they require significant development effort and expertise in the form of extensive manual optimization of datapaths, pipelines, and memory layouts \cite{liao2024system}. HLS, in contrast, provides faster development cycles and improved portability, but it faces several challenges specific to cryptographic applications, including (i) ensuring constant-time operation to mitigate timing side-channel vulnerabilities, (ii) efficiently mapping modular arithmetic and large polynomial operations to FPGA resources, (iii) managing memory access patterns to minimize bottlenecks in polynomial transforms, and (iv) supporting flexibility across different security levels and parameter sets \cite{gleissenthall2019iodine, lu2024efficient}.

Recent advances in Large Language Models (LLMs) offer new opportunities to address these challenges. LLMs can parse algorithmic descriptions, generate synthesizable hardware descriptions, and suggest optimizations by recognizing patterns across diverse hardware designs \cite{liao2024llmsgoodhighlevelsynthesis}. Early demonstrations of LLMs in hardware design tasks have shown promise in generating HDL and assisting HLS flows \cite{thakur2023verigenlargelanguagemodel,collini2024c2hlscllmsbridgesoftwaretohardware}, but their application to cryptographic hardware acceleration remains unexplored. Unlike conventional HLS tools, which rely on predefined pragmas, LLMs can interpret algorithmic semantics and propose hardware-level optimizations without explicit directives, bridging the gap between algorithm understanding and architecture generation. PQC introduces unique requirements---correctness under strict modular arithmetic, constant-time execution for security, and scalability across parameter sets---that make it a compelling, yet demanding target for LLM-driven design automation.

In this work, we explore the intersection of HLS and LLM capabilities to accelerate PQC hardware design. Specifically, we propose a methodology where LLMs assist in both hardware-software partitioning and HDL generation, reducing design effort while maintaining cryptographic correctness and efficiency. To ground this methodology, we use FALCON \cite{FALCON2025} as a case study, demonstrating the potential of LLMs to complement traditional HLS flows and enable rapid development of FPGA-based PQC accelerators.

\section{Methodology}
Fig. \ref{fig:methodology} illustrates the proposed framework for LLM-driven hardware acceleration of post-quantum cryptography. The process begins with a C/C++ implementation of a PQC algorithm, which serves as the input for either traditional profiling-based partitioning or LLM-driven hardware–software partitioning. Performance-critical components identified in this stage are then passed to the hardware generation phase, where they can be synthesized using conventional high-level synthesis (HLS) tools or generated directly by LLMs through tailored prompts. The resulting hardware modules are integrated with software (e.g., ARM-based), forming a complete hardware–software system. Finally, the generated designs undergo verification, synthesis, and implementation using FPGA toolchains to evaluate performance, area, and power trade-offs.  

This framework highlights two key opportunities where LLMs can enhance the PQC design workflow: (i) assisting in hardware-software partitioning by analyzing algorithmic structure and profiling data, and (ii) generating candidate hardware descriptions and constraints for FPGA synthesis. In the following subsections, we describe these components in detail.

\subsection{Hardware-software partitioning}\label{Sec:SH_partition}
Hardware-software co-design is a systematic methodology for jointly optimizing hardware and software components of embedded systems to meet performance, power, and cost constraints \cite{HWSWcodesign_past_present_future}. At its core, hardware-software partitioning determines which system components should be implemented in hardware versus software. Classical approaches rely on formal system models such as Data Flow Graphs (DFGs), Control Data Flow Graphs (CDFGs), or Directed Acyclic Graphs (DAGs), which enable rigorous analysis of computational patterns, data dependencies, and resource requirements.

However, PQC algorithms present unique challenges for partitioning. Unlike traditional embedded workloads, PQC schemes are primarily distributed as reference software implementations (in C, C++, and Python) through the NIST standardization process, without corresponding architectural specifications or formal computation models. This creates a fundamental gap: traditional co-design methodologies assume the availability of formal system models, while PQC development is rooted in software-centric distributions. As a result, designers must bridge the gap between algorithm-level descriptions and hardware-amenable computational kernels, particularly for operations such as polynomial arithmetic, transforms, and Gaussian sampling.

Traditional approaches to identifying hardware acceleration candidates in software implementations rely on performance profiling tools like GNU gprof \cite{graham1982gprof} to analyze the execution patterns, resource utilization, and data dependencies. This empirical analysis helps identify computationally intensive functions and operations that could benefit from hardware acceleration. While effective, this process requires expert interpretation of profiling data, deep knowledge of both cryptography and hardware design, and careful reasoning about which functions merit acceleration. This expertise barrier makes partitioning a bottleneck in PQC accelerator development. 

To address this challenge, we propose raising the abstraction level of partitioning with LLMs. LLMs can process profiling data or directly analyze source code, understand algorithmic structures, and recommend candidate functions for hardware acceleration. As illustrated in Figure \ref{fig:methodology}, our methodology supports two parallel paths:
\begin{enumerate}
    \item \textbf{Profiler-guided partitioning:} profiling tools (e.g., gprof) produce performance traces, which are then summarized and analyzed by LLMs to recommend hardware acceleration targets.
    \item \textbf{Source-guided partitioning:} LLMs analyze PQC source code directly, without profiling, to infer computational hotspots and propose acceleration candidates.
\end{enumerate}

Both paths converge on the identification of functions most amenable to FPGA-based acceleration. These recommendations are subsequently passed to either conventional HLS tools or LLM-driven HDL generation, ultimately producing synthesizable hardware descriptions that can be verified using industry-standard tools like Vivado, ModelSim, or VCS. 

A key practical consideration in the LLM-based approach is managing input granularity. In our experiments, we evaluate two strategies for source-guided partitioning: (i) \textbf{abstract prompts}, where only top-level function names or algorithm identifiers (e.g., "FALCON signature generation") are provided, testing the LLM's ability to reason from its pre-trained knowledge; and (ii) \textbf{full-code prompts}, where the complete implementation is provided, enabling fine-grained reasoning about computational bottlenecks. Figure \ref{fig:LLM_ranking} illustrates example prompts used in these workflows.

Finally, token-context limitations and prompt-engineering strategies play an important role in scaling LLM partitioning to complex PQC algorithms. Algorithms such as FALCON, which involve deeply nested polynomial arithmetic routines, require careful segmentation of code and prompts to ensure accurate analysis. In Section \ref{sec:experiments}, we quantitatively compare the accuracy of LLM-generated recommendations against GNU gprof profiling and present a case study on FALCON to evaluate how well LLMs can identify performance-critical kernels across both partitioning paths.

\newsavebox{\codeboxA}
\newsavebox{\codeboxB}

\begin{lrbox}{\codeboxA}
\begin{minipage}{0.95\linewidth}
{\scriptsize
\begin{lstlisting}
Program: Post-quantum cryptography, FALCON
Source code: provided reference implementation
Task: Perform a static cost analysis of the key-generation path. Based on algorithmic complexity, hot loops, and heavy-number routines, estimate which functions will consume the most CPU time
\end{lstlisting}
}
\end{minipage}
\end{lrbox}

\begin{lrbox}{\codeboxB}
\begin{minipage}{0.95\linewidth}
{\scriptsize
\begin{lstlisting}
1.  zint_add_scaled_mul_small  (*@O($N^2$)@*)
2.  zint_add_mul_small  O($N^2$)
3.  poly_inv  O($N^2$)
4.  modp_montymul  O(1) per call but called O($N^2$)
Takeaways for Acceleration: target the two zint routines (zint_add_scaled_mul_small and zint_add_mul_small) first, they soak up the lion$'$s share of time. Offload modp_montymul into a vectorized or hardware-accelerated block to accelerate everything that calls it.
\end{lstlisting}
}
\end{minipage}
\end{lrbox}

\begin{figure}
\centering
\subfloat[Prompt\label{fig:prompt_ranking}]{\usebox{\codeboxA}}

\vspace{-5pt}

\subfloat[Feedback\label{fig:feedback_ranking}]{\usebox{\codeboxB}}

\caption{Using LLM to rank functions in the PQC's key generation path.}
\vspace{5pt}
\label{fig:LLM_ranking}
\end{figure}

\newsavebox{\codeboxC}
\newsavebox{\codeboxD}

\begin{lrbox}{\codeboxC}
\begin{minipage}{0.95\linewidth}
{\scriptsize
\begin{lstlisting}[]
Source code: modp$\_$montymul function
Task: Implement the function in fully pipelined Verilog for the Zynq UltraScale+ MPSoC ZCU104 board, utilizing DSP48E2 primitives.
\end{lstlisting}
}
\end{minipage}
\end{lrbox}

\begin{lrbox}{\codeboxD}
\begin{minipage}{0.95\linewidth}
{\scriptsize
\begin{lstlisting}[]
...
module modp_montymul_pipelined (
    input  wire        clk,      // Clock signal
    input  wire        rst,      // Reset 
    input  wire        start,    // Valid input
    input  wire [31:0] a,        // 32-bit operand A
...
always @(posedge clk) begin
        if (rst) begin
            // Clear all pipeline registers
            a_s0      <= 32$'$d0; b_s0    <= 32$'$d0;
            p_s0      <= 32$'$d0; p0i_s0  <= 32$'$d0;
            v_s0      <= 1$'$b0;
...
        end else begin
            // Stage 0 -> Stage 1
            a_s0      <= a;
            b_s0      <= b;
            p_s0      <= p;
            p0i_s0    <= p0i;
            v_s0      <= start;
...

\end{lstlisting}
}
\end{minipage}
\end{lrbox}

\begin{figure}
\centering
\subfloat[Prompt\label{fig:LLM_generated_input}]{\usebox{\codeboxC}}

\vspace{-5pt}

\subfloat[Feedback\label{fig:LLM_generation_feedback}]{\usebox{\codeboxD}}

\caption{Using LLM to accelerate modp\_montymul function with fully pipelined design.}
\label{fig:LLM_generation}
\end{figure}

\subsection{LLM-driven hardware generation}
Once hardware-software partitioning identifies acceleration candidates, the next challenge is translating these algorithmic components into efficient hardware implementations. Traditional hardware development for PQC algorithms has relied heavily on HLS approaches \cite{HLSNISTPQC2019}. While HLS tools reduce development time compared to manual RTL design, they often produce suboptimal architectures that do not exploit algorithm-specific optimization opportunities. Our LLM-driven approach focuses on translating the partitioning outcomes into synthesizable hardware descriptions in Verilog, emphasizing pipeline structure, resource mapping, and interface generation.

The generation of synthesizable hardware accelerators for computationally intensive PQC operations requires careful prompt engineering to guide LLMs in producing optimized RTL implementations. For FALCON's core operations---such as its Number Theoretic Transform (NTT) and NTRU polynomial arithmetic---we developed a structured prompting framework that encompasses both architectural and implementation-specific requirements. This framework begins by establishing the hardware design context, including target device specifications (e.g., Xilinx xczu7ev-ffvc1156-2-e FPGA), performance objectives, and resource constraints. The prompt structure ensures the generated Verilog code not only meets functional requirements but also adheres to synthesizable RTL coding guidelines and cryptographic hardware security principles. Figure \ref{fig:LLM_generation} shows an example of using an LLM to generate Verilog code for modp\_montymul, the Montgomery modular‐multiplication function, with a fully pipelined design that utilizes DSP48E2 primitives. In practice, generating high-quality accelerators requires iterative prompt refinement. For the kernels studied in this work, each Verilog module typically required approximately \textit{15–20} LLM–tool iterations, including initial code generation, syntax correction, functional validation, and timing refinement. We later use this iteration count as a coarse proxy for quantifying human effort in the LLM-driven flow.

When generating RTL for computationally intensive functions like FALCON's NTT operations, the algorithm's complexity may exceed typical LLM token limits (e.g., 128k tokens for GPT-4). We address this through hierarchical decomposition, breaking down complex modules into logically coherent subsystems \cite{duong2021_butterfly}, modular arithmetic cores, and memory interfaces, each of which can be generated and validated independently. Post-generation validation encompasses multiple levels: syntactic correctness, adherence to RTL coding guidelines, functional verification against reference software implementations, and validating functional correctness while respecting cryptographic design constraints such as fixed-latency execution. This LLM-driven approach enables iterative refinement of the generated RTL through natural language feedback, leveraging the LLM's understanding of both hardware design and cryptographic requirements to improve functionality and optimization.

The final phase of the hardware generation process involves a comprehensive integration and verification infrastructure. We generate Tool Command Language (TCL) scripts for automated IP core packaging and integration into the target FPGA design flow, along with physical design constraints specified in Xilinx Design Constraint (XDC) files for timing, placement, and I/O requirements. The testbench generation focuses on validating critical PQC operations under various test vectors, including corner cases specific to lattice-based cryptography such as polynomial coefficient overflow conditions and modular reduction edge cases. While LLMs significantly accelerate the development process by automating RTL generation and refinement, successful deployment of PQC hardware accelerators ultimately requires careful consideration of multiple factors, including cryptographic correctness guarantees, side-channel resistance, performance requirements across different security levels, and compliance with emerging PQC standards. The iterative nature of our LLM-driven approach, combined with domain expertise in both hardware architecture and cryptographic implementation security, enables the systematic development of efficient and secure PQC accelerators.

\begin{table}[t]
    \caption{Hardware/Software partitioning for the FALCON key-generation algorithm using ChatGPT-4o and GNU gprof profiling, with the software compiled using -O3}
    \vspace{-5pt}
    \label{tab:Ranking_top}
    \centering
    \resizebox{0.46\textwidth}{!}{
    \begin{tabular}{|c|c|c|}
    \hline
    \multicolumn{2}{|c|}{\textbf{GNU gprof}} & \textbf{ChatGPT-4o} \\
    \hline
    \textbf{Function name} & \textbf{\% Runtime} & \textbf{Function name}\\ 
    \hline
    solve\_NTRU\_intermediate & 32.76 & solve\_NTRU\\ 
    zint\_rebuild\_CRT & 21.84 & Zf(sign\_dyn)\\ 
    falcon\_inner\_keygen & 12.97 & Zf(sampler)\\ 
    poly\_small\_mkgauss & 7.85 & Zf(FFT) / Zf(iFFT)\\ 
    process\_block & 7.51 & poly\_small\_mkgauss\\ 
    \hline
    \end{tabular}
    }
    \vspace{-5pt}
\end{table}

\begin{table}[t]
    \caption{Hardware/Software partition for the FALCON key-generation algorithm using ChatGPT-4o and GNU gprof profiling, with software compiled with -O3 -fno-inline}
    \vspace{-5pt}
    \label{tab:Ranking_low}
    \centering
    \resizebox{0.46\textwidth}{!}{
    \begin{tabular}{|c|c|c|}
    \hline
    \multicolumn{2}{|c|}{\textbf{GNU gprof}} & \textbf{ChatGPT-4o} \\
    \hline
    \textbf{Function name} & \textbf{\% Runtime} & \textbf{Function name}\\ 
    \hline
    modp\_montymul & 18.52 & zint\_add\_scaled\_mul\_small\\ 
    modp\_add & 18.51 & zint\_add\_mul\_small\\ 
    zint\_add\_scaled\_mul\_small & 14.81 & modp\_montymul\\ 
    \_init & 6.17 & poly\_inv\\ 
    zint\_mod\_small\_unsigned & 4.95 & poly\_mul\_ntt\\ 
    \hline
    \end{tabular}
    }
\end{table}

\begin{table*}[t]
    \caption{Place-and-route results using ChatGPT-4o and Vitis HLS generated hardware accelerators for low-level compute-intensive functions}
    \vspace{-5pt}
    \label{tab:PRresult}
    \centering
    \resizebox{0.9\textwidth}{!}{
    \begin{tabular}{cccccccccc}
    \hline
    \textbf{Function} & \textbf{Baseline (ns)} & \textbf{Approach} & \textbf{CP (ns)} & \textbf{Cycles} & \textbf{Time (ns)}& \textbf{LUT}  & \textbf{FF} & \textbf{DSP} & \textbf{Power (W)} \\ 
    \hline
    \multirow{2}{*}{modp\_montymul} & \multirow{2}{*}{45.73} & LLM & 4.805 & 6 & 28.83 & 267 & 443 & 11& 0.725\\ 
    & & HLS & 9.646 & 3 & 28.94 & 247 & 246 & 11 & 0.775\\ 
    \hline
    \multirow{2}{*}{zint\_add\_scaled\_mul\_small} & \multirow{2}{*}{1288.43} & LLM & 6.848 & 101 & 691.65 & 195 & 161 & 3& 0.748\\ 
    & & HLS & 8.955 & 199 & 1782.05 & 548 & 231 & 4 & 0.698\\ 
    \hline
    \multirow{2}{*}{modp\_add} & \multirow{2}{*}{25.26} & LLM & 3.715 & 3 & 11.15 & 80 & 228 & 0 & 0.703\\ 
    & & HLS & 5.36 & 4 & 21.44 & 94 & 102 & 0& 0.759\\ 
    \hline
    \multirow{2}{*}{zint\_mod\_small\_unsigned} & \multirow{2}{*}{426.31} & LLM & 4.503 & 67 & 301.70 & 451 & 750 & 11& 0.627\\ 
    & & HLS & 12.681 & 30 & 380.43 & 538 & 310 & 11& 0.671\\ 
    \hline
    \end{tabular}
    }
    \vspace{-10pt}
\end{table*}

\section{Experiments}\label{sec:experiments}

\subsection{Experimental setup}
To evaluate the effectiveness of the proposed framework, we selected FALCON (Fast-Fourier Lattice-based Compact Signatures over NTRU) as a representative case study of lattice-based post-quantum digital signature schemes. Our experiments began with the official C reference implementation of FALCON, which we compiled using GCC and profiled on an ARM Cortex-A53 processor to establish baseline performance. Profiling was performed with GNU gprof to identify computational hotspots and guide hardware-software partitioning. For LLM-based workflows, we used the ChatGPT-4o model to analyze profiling results and source code, as well as to generate candidate hardware descriptions. Specifically, the LLM is prompted to produce Verilog modules, testbenches, TCL scripts, and XDC constraint files for FPGA synthesis.

For hardware development, we targeted the AMD Zynq UltraScale+ MPSoC ZCU104 development board, which integrates a quad-core Cortex A53 and programmable logic fabric, enabling full hardware-software co-execution of PQC workloads. HLS development is performed using Vitis HLS 2023.2. Both design flows proceed through verification, place-and-route, and implementation using the Xilinx Vivado toolchain. Each generated module was functionally simulated in Vivado using LLM-generated self-checking testbenches, followed by synthesis and post-place-and-route timing analysis to verify correctness.

\subsection{Results and analysis}
We evaluate performance at the kernel level after place-and-route to isolate the architectural impact of the proposed acceleration. Such kernel-level characterization provides a necessary foundation for understanding system behavior, while overall end-to-end performance may additionally depend on factors including memory movement, PS–PL communication, and control overhead.

\subsubsection{Hardware/software partitioning}
We first evaluated the hardware/software partitioning capabilities of ChatGPT-4o by comparing its analysis against the GNU profiler (gprof). Using ChatGPT-4o, we analyzed three main functionalities---\textit{key generation, signature generation, and signature verification}---by inputting their top-level functions into ChatGPT-4o and prompting it to identify the most computationally intensive function. The LLM indicated that key generation was the most dominant compute-intensive operation among the three.

To validate this prediction, we used gprof to profile the runtime of all three functionalities. The profiling results confirmed that key generation dominated the runtime, leading us to conduct a deeper functional partitioning analysis of the key generation algorithm. To account for compiler-level optimizations that influence profiling visibility, we evaluated both inline-enabled and inline-disabled builds. With inlining enabled, the compiler merges frequently used routines into higher-level functions, providing a realistic view of actual runtime behavior. However, this optimization also hides fine-grained kernel boundaries, making it difficult to identify candidate functions for hardware offloading. Therefore, we also profiled the implementation with \texttt{-fno-inline}, which exposes individual routines and enables more precise kernel-level analysis for the LLM. Table \ref{tab:Ranking_top} compares the functions identified by gprof and ChatGPT-4o under the standard \texttt{-O3} optimization flag. The gprof results show that \texttt{solve\_NTRU\_intermediate} (32.76\% runtime), \texttt{zint\_rebuild\_CRT} (21.84\%), and \texttt{falcon\_inner\_keygen} (12.97\%) dominate the runtime distribution. On the other hand, ChatGPT-4o identified functions like \texttt{solve\_NTRU}, \texttt{Zf(sign\_dyn)}, and the Gaussian sampler (\texttt{Zf(sampler)}) as critical. 

To further evaluate ChatGPT-4o, we recompiled FALCON with \texttt{-O3 -fno-inline} optimization flag to expose low-level routines that are otherwise inlined. The gprof profiling revealed that \texttt{modp\_}\texttt{montymul} (Montgomery multiplication) had the highest call frequency and runtime contribution. ChatGPT-4o also identified \texttt{modp\_}\texttt{montymul} among its top three hotspots, demonstrating alignment with empirical profiling at the fine-grained level (Table \ref{tab:Ranking_low}). Minor deviations in intermediate function rankings suggest that ChatGPT-4o, when analyzing only source code, was less sensitive to control flow complexity and runtime call frequency compared to dynamic profiling. Overall, these results indicate that LLM-based hardware–software analysis can provide useful high-level insight into program structure and often highlight computationally expensive kernels such as NTRU solvers, Gaussian samplers, and modular multipliers. However, we also observed noticeable deviations in intermediate function rankings, which confirms that profiling tools remain indispensable for accurately quantifying runtime contributions. LLMs are best positioned as decision-support tools and can narrow down candidate functions for acceleration, while profilers provide the runtime validation needed for final hardware–software partitioning decisions.

\subsubsection{Accelerator generation and comparison to prior work}
Based on the results from ChatGPT-4o and gprof, we generated hardware accelerators for the most compute-intensive low-level functions identified in Table \ref{tab:Ranking_low}, excluding initialization routines (\texttt{\_init}). We compared accelerators generated by ChatGPT-4o with those produced by Vitis, following methodologies similar to prior FPGA PQC studies \cite{nguyen2020high}. Table \ref{tab:PRresult} reports the 10-run average execution time of each function on the Cortex-A53 baseline, along with post place-and-route results (resource usage, execution time, and power consumption) for both LLM- and HLS-generated accelerators. 

Overall, both flows achieved performance improvements over the software baseline, with average speedups of 1.782× for LLM-generated accelerators and 1.15× for HLS-generated accelerators. The higher speedup of the LLM-based approach is attributable to two specific optimizations: (i) we aggressively pipelined all functions to reduce latency and critical-path delay, and (ii) we mapped multiplications to DSP48E2 slices to shorten multiplier delays. For example, in \texttt{modp\_montymul}, the LLM design achieved a critical path of 4.805 ns versus 9.646 ns for the HLS design, halving the delay through deeper pipelining.

In terms of energy and resource trade-offs, the results indicate complementary strengths of the two flows. Across the functions (excluding  \texttt{zint\_add\_scaled\_mul\_small}), the LLM-generated accelerators reduced power consumption by 6.99\% relative to HLS designs. However, this came at the expense of higher resource usage: on average LLM designs consumed 31.86\% more LUTs and FFs. For example, \texttt{modp\_montymul} required 443 flip-flops in the LLM design compared to 246 in HLS, reflecting the deeper pipelining strategies. Interestingly, despite the larger register count, dynamic power was lower in the LLM design possibly due to reduced combinational glitching and decreased switching activity per pipeline stage.

One outlier is \texttt{zint\_add\_scaled\_mul\_small}, where the HLS-generated accelerator underperformed the software baseline, achieving only 0.723× the speed of the Cortex-A53 implementation. This slowdown is likely due to Vitis HLS not applying pipelining directives, resulting in higher latency and a longer critical path. In contrast, the LLM-generated design achieved a 1.86x speedup for the same function.

Taken together, these results demonstrate that LLM-based accelerator generation can produce designs with competitive or superior performance to conventional HLS, especially when exploiting pipeline parallelism. However, they also reveal a clear power–resource trade-off: LLM-generated accelerators tend to consume more LUTs and FFs, while HLS designs are more resource-efficient. Moreover, we observed that effective exploitation of LLMs still requires domain expertise, for example, in crafting prompts that highlight PQC-specific operations, interpreting trade-offs between latency, power, and area, and validating correctness under cryptographic constraints. This suggests that future flows may benefit from hybrid approaches that involve using LLMs for rapid exploration and aggressive optimization, while relying on expert-guided HLS refinement for area efficiency and security assurance.

\subsubsection{Security and Constant-Time Considerations}
Constant-time execution is a critical requirement for deployed cryptographic accelerators to mitigate timing side-channel attacks. In this work, constant-time behavior is treated as a design intent rather than a formally verified property. Specifically, LLM-driven hardware generation is guided to produce fixed-latency pipelines and to avoid data-dependent control flow and memory access patterns. However, this study does not perform formal constant-time verification or side-channel leakage analysis. Such verification typically requires specialized formal methods or power and timing measurements and is orthogonal to the performance-oriented focus of this work. We leave rigorous constant-time verification to future work.

\section{Conclusion}
The rapid advancement of quantum computing underscores the urgent need for robust post-quantum cryptographic (PQC) solutions. This paper explored the use of Large Language Models (LLMs), specifically ChatGPT-4o, to accelerate hardware-software co-design for PQC, with FALCON as a case study. Our results show that LLMs can achieve competitive performance relative to conventional HLS. At the same time, effective use of LLMs still requires domain expertise to guide prompt design, interpret trade-offs, and ensure functional correctness under cryptographic constraints. Moreover, critical challenges remain around addressing constant-time execution requirements, avoiding inadvertent vulnerabilities, and establishing rigorous verification flows. Overall, while not a replacement for human expertise, LLMs represent a promising assistive technology for rapid, adaptive PQC accelerator design on FPGAs. 

Although this study evaluates individual kernels in isolation, the proposed framework is system-aware by construction, as it explicitly targets HW/SW partitioning and PS–PL integration. Future work will extend this framework toward full system-level acceleration by integrating multiple kernels and evaluating end-to-end PS–PL execution on the ZCU104 FPGA. While FALCON is used as a representative case study, the proposed profiling, partitioning, and accelerator-generation flow is not specific to a single algorithm. We plan to apply the methodology to additional standardized lattice-based PQC schemes, such as CRYSTALS-Kyber and Dilithium, which share similar polynomial arithmetic and modular reduction structures, to further validate the generality and robustness of the LLM-driven HW/SW co-design approach.

\balance
\bibliographystyle{IEEEtran}
\bibliography{References}

\end{document}